
\magnification\magstep1


\def\R{{\bf R}}
\def\S{{\cal S}}
\headline={\ifnum\pageno=1 {\tenrm gr-qc/9303011\hfil} \else\hfil\fi}

\centerline{\bf THE CONSTRUCTION OF SORKIN TRIANGULATIONS}
\bigskip
\centerline{Philip A.~Tuckey\footnote{$^{\rm\dag}$}{E-mail:
pat@lavu.physics.lancaster.ac.uk}}
\smallskip
\centerline{School of Physics and Materials, Lancaster University}
\centerline{Lancaster, LA1 4YB, United Kingdom}

\bigskip
{\narrower
Some time ago, Sorkin (1975) reported investigations of the time evolution
and initial value problems in Regge calculus, for one triangulation each of
the manifolds $\R\times{\rm S}^3$ and $\R^4$. Here we display the simple,
local characteristic of those triangulations which underlies the structure
found by Sorkin, and emphasise its general applicability, and therefore the
general validity of Sorkin's conclusions.  We also make some elementary
observations on the resulting structure of the time evolution and initial
value problems in Regge calculus, and add some comments and speculations.
\par}

\bigskip\bigskip
In Regge calculus (Regge 1961, see Williams and Tuckey 1992 for a brief
review), the 4-dimensional spacetime manifold $M$ is triangulated,
i.e.~divided into cells which are 4-simplices, and the metric is
approximated as being flat on the interiors of these cells. The metric is
determined by the squared lengths $s_\lambda$ of the 1-simplices
(``edges'') $\lambda$ in the triangulation, and curvature is concentrated
on the 2-simplices (``bones''). The Einstein-Hilbert action $\int
R\sqrt{|g|}{\rm d}^4 x$ can be evaluated for these piecewise-flat metrics,
and leads to
$$
S = \sum_{{\rm bones}\ b} \varepsilon_b A_b \ ,
\eqno (1)$$
where $A_b$ is the area of bone $b$, and $\varepsilon_b$ is its deficit
angle, defined as follows.

Let $\alpha$ be a ($0,1,2,3 \hbox{ or } 4$-)simplex. Define the ``$n$-star of
$\alpha$'', $\S_n(\alpha)$, to be the set of all $n$-simplices which either
contain or are contained in $\alpha$. (The extension to the $n$-star of a set
of simplices is straightforward.) The deficit angle of bone $b$ is
$$
\varepsilon_b = 2\pi - \sum_{c\in\S_4(b)}\theta_{bc} \ ,
\eqno (2)$$
where $\theta_{bc}$ is the dihedral angle between the two tetrahedral faces of
cell $c$ which meet at bone $b$, measured internally to $c$.

The discrete ``equations of motion'', approximating Einstein's equations,
are the Regge equations,
$$
{\partial S \over \partial s_\lambda} = 0
\ \ \ \ \forall\hbox{ edges }\lambda \ ,
\eqno (3)$$
i.e.~one algebraic equation for each edge $\lambda$. It was proved by Regge
(1961) and Sorkin (1975) that the terms resulting from the variations of the
deficit angles in (1) cancel, so equations (3) become just
$$
\sum_{{\rm bones}\ b} \varepsilon_b {\partial A_b \over \partial s_\lambda}
= 0 \ \ \ \ \forall\ \lambda \ .
\eqno (4)$$
It follows that the equation of motion for edge $\lambda$ involves only
$\S_4(\lambda)$ (Sorkin 1975). Following Sorkin, if edge
$\mu\in\S_1(S_4(\lambda))$, we say that ``$\lambda$ implicates $\mu$''.

In the two cases he reported, Sorkin (1975) found that for the time
evolution problem the Regge equations separated into small sets, which
could be solved independently to evolve a local region of a spacelike
hypersurface forward in time. This structure is much simpler than in the
``3+1'' and ``null-strut'' approaches developed later (Miller 1986,
Porter 1987, Tuckey 1989), where the evolution equations
across a whole hypersurface are coupled, giving a massive algebraic problem
to solve at each step in a time evolution calculation.  We believe that the
reason that Sorkin's results have not attracted more attention is that the
simple property of his triangulations which underlies them was not
generally understood, so the general validity of his findings was not
appreciated. We hope to correct this situation with this note.

Sorkin also identified a sufficient amount of initial data needed to allow
a time evolution calculation to commence, as well as the Regge equations
which concern only these data, which together constitute the initial value
problem in Regge calculus.  Our emphasis on the basic structure of the
triangulations used also allows an interesting further analysis of this
problem.

\smallskip
\leftline{\it Time Evolution Problem}

\smallskip
We assume $M$ has topology $\R\times\Sigma$, where $\Sigma$ is a 3-manifold,
taken to have no boundary for convenience. To display the typical step in
a time evolution calculation we assume we have initial data on a slice of $M$,
which is bounded by 3-dimensional hypersurfaces $\Sigma_1$ and $\Sigma_2$
having the topology of $\Sigma$, and which divides $M$ into two separate
regions, denoted as  ``past'' and ``future''. More specifically, we assume we
have some triangulation of this 4-dimensional slice (which induces
triangulations of $\Sigma_1$ and $\Sigma_2$), and that the squared lengths of
all edges in this triangulated region are known. We further require that no
edge should lie in both $\Sigma_1$ and $\Sigma_2$. Thus $\Sigma_1$ and
$\Sigma_2$ can intersect at vertices of their triangulations, but not on
higher-dimensional simplices.

The typical time evolution step proceeds by extending the triangulation into
the future region of $M$, from a local region of the future hypersurface, say
$\Sigma_2$. This is done by choosing some vertex $\lambda$ on $\Sigma_2$,
adding a new vertex $\lambda'$ in the future region of $M$, and then attaching
a new 4-simplex to each tetrahedron in $\Sigma_2$ which contains $\lambda$,
with all of these new 4-simplices sharing $\lambda'$ as their one vertex which
is not in $\Sigma_2$. The figure shows the 3-dimensional analogue of this
procedure.

Denote by $\mu_1,\ldots,\mu_{N_1}$ the elements of $\S_0(S_1(\lambda))$
which lie in $\Sigma_2$, not including $\lambda$. $N_1$ is the number of
edges on $\Sigma_2$ which intersect at $\lambda$. Our construction has
introduced $N_1+1$ new edges; the $N_1$ ``diagonal'' edges $\mu_1\lambda'$,
$\mu_2\lambda'$,\dots,$\mu_{N_1}\lambda'$, and the ``vertical'' edge
$\lambda\lambda'$. We have also made $N_1+1$ new edges lie in the interior
of the triangulated region, the $N_1$ ``horizontal'' edges $\mu_1\lambda$,
$\mu_2\lambda$,\dots,$\mu_{N_1}\lambda$, as well as $\lambda\lambda'$. The
equations of motion for these new interior edges implicate the $N_1+1$ new
edges, plus edges lying in the initial slice whose squared lengths are
already known. Thus, in principle, we can solve these equations to find the
squared lengths of the new edges.

We note that the extended, triangulated slice of $M$ thus obtained
satisfies the conditions required of the initial data, so the process may
be repeated again at any vertex on the new future boundary hypersurface.
Clearly this construction is purely local, and may be carried out regardles
of the topology of $\Sigma$, or the triangulation of $\Sigma_2$. Thus this
construction allows the time evolution problem in Regge calculus to consist
in general of the successive solution of sets of $N_1+1$ equations in
$N_1+1$ unknowns, where $N_1$ is the number of edges in the current
hypersurface meeting on the relevant vertex.

We refer to this basic evolution step as ``the evolution of a vertex'', and
call a triangulation built up from a suitable initial slice by a series of
such steps a ``Sorkin triangulation''. We note that there is a one to one
correspondence between the horizontal and diagonal edges referred to above,
and indeed between all simplices in the original hypersurface $\Sigma_2$
and those in the new future boundary of the triangulated region. In fact,
the evolution of a vertex leaves invariant both the topology and
triangulation of the spacelike hypersurfaces. Thus a Sorkin triangulation
may be viewed as being based on a sequence of identically triangulated
3-dimensional hypersurfaces, which overlap each other in some regions and
which are separated by vertical edges (and associated 4-simplices) in
others.

Presumably one would work systematically across a hypersurface, evolving
all vertices once before any is evolved twice, although one might choose
not to evolve some region to avoid a singularity, for example. It is worth
noting that any two vertices on a hypersurface which are not connected by
an edge may be evolved independently and hence simultaneously, as the
corresponding sets of equations are not coupled. Thus the method lends
itself to parallel processing.

For example, if $\Sigma$ is ${\rm S}^3$ and the triangulation of a given
hypersurface is the 600-tetrahedron regular triangulation, having 120
vertices and 12 edges meeting on each vertex, then all
vertices could be evolved forward once by a process of 4 steps, in each of
which 30 are evolved in parallel (Williams 1990). The evolution of each
vertex would involve the solution of 13 equations in 13 unknowns.

\smallskip
\leftline{\it Initial Value Problem}

\smallskip
The initial value problem, in short, is to ensure that the initial data,
i.e.~the squared lengths of the edges in the initial slice, satisfy the
relevant Regge equations. The relevant equations are those for edges lying in
the interior of the initial slice, since these implicate only edges in the
initial slice. (The equations for edges lying on the boundary of the slice
are relevant to the evolution of the metric into the future or past regions of
$M$, as described above.)

The initial value problem may be viewed as a sequence of incomplete time
evolution steps in the following way. We commence with the 3-dimensional
hypersurface $\Sigma_1$, triangulated in some way, and we freely specify
the squared lengths of the edges in this triangulation. Then we attempt to
evolve a vertex on $\Sigma_1$ in the way described above. This introduces
only one new edge on the interior of the triangulated region, the vertical
edge, so the $N_1+1$ new edge lengths may be chosen subject only to the
equation corresponding to this edge. (Of course, the simplex inequalities
which ensure that the metric is Minkowskian on the interiors of the
4-simplices must also be respected.)

Many vertices on $\Sigma_1$ may be evolved in this way, but eventually we
reach a stage where the evolution of a vertex involves adding 4-simplices
onto 4-simplices which have previously been added to $\Sigma_1$. At this stage
we will have to satisfy both the equation for the new vertical edge, and those
for any horizontal edges which are now put into the interior of the
triangulated region. We continue making such partial evolution steps until any
further evolution would involve a full set of $N_1+1$ equations, when the
initial data is complete. Thus the initial value problem may be viewed as the
solution of a sequence of sets of less than $N_1+1$ equations in $N_1+1$
variables.

Note that any further evolution step involves a full complement of
equations when the initial slice has reached a state where no edge in the
future boundary hypersurface also lies in $\Sigma_1$, which gives the
condition we required of the initial data earlier. For example, taking the
600-tetrahedron triangulation of ${\rm S}^3$ mentioned above, this could be
achieved by making 3 steps, in each of which 30 vertices are evolved,
leaving 30 vertices on $\Sigma_1$ unevolved. These considerations are
consistent with the initial data identified by Sorkin (1975) in his
examples, but the analysis given here is new.

\smallskip
\leftline{\it Further Comments}

\smallskip
We have not quantitatively addressed the crucial question of the existence
of solutions, but we speculate that relativistic causality is relevant
here. We expect that in the evolution of any vertex $\lambda$ the diagonal
edges will typically be spacelike or null, since we expect the new
4-simplices added to lie within the future light cone of the known part of
$\S_4(\lambda)$, which contains all the initial data for the evolution
step. The vertical edge may be timelike, null or spacelike, but the
condition on the diagonal edges implies a restriction on this edge,
corresponding to a Courant limit. We thus expect the family of
3-dimensional hypersurfaces referred to above to be spacelike (or null)
typically, while the edges going between them can be timelike, spacelike or
null.

In making a time evolution calculation, it is thought (Tuckey 1991, Galassi
and Miller 1992) that it will be appropriate to exploit the approximate
symmetries of Regge calculus (Ro\v cek and Williams 1984, Hartle 1985, Piran
and Strominger 1986)  to allow some of the new, unknown edge lengths to be
freely specified, and to ignore some of the equations of motion (or rather to
use them as a check on the solutions). Important work on this point has been
carried out by Galassi and Miller (1992), both in demonstrating numerically
the effects of the approximate symmetries on the Regge equations, and in
identifying combinations of edge lengths to treat as ``gauge variables''.
This approach corresponds to freely specifying the discretised lapse and
shift in finite-difference approximations to Einstein's equations, and using
the corresponding constraint equations only to check the solutions, even
though the diffeomorphism symmetry of the continuum theory is broken in this
approximation. In the continuum limit the diffeomeorphism symmetry should be
restored (in both approximations), so in this limit the solutions will
satisfy all equations.

This approach is in contrast with that envisaged by Sorkin, who wished to
solve all equations at each time evolution step to find all new edge lengths,
thus ``allowing the equations to choose the gauge''. It is expected that
this approach would lead to ill-conditioned algebraic systems, because of the
approximate freedom in the solutions, as found in previous numerical work
(Hartle 1986).

Barrett (1992) has given a beautiful interpretation of the basic evolution
step described here as a sequence of elementary moves on the triangulation of
the spacelike hypersurface, satisfying a ``regularity condition'': that the
number of new edges introduced is equal to the number of new edges on the
interior of the triangulated region. From this point of view we may state an
important question: are there sequences of elementary moves which satisfy the
regularity condition, but which change the triangulation of the spacelike
hypersurfaces? Such sequences would provide a way of adaptively refining the
triangulation of space, allowing the insertion or deletion of simplices as
appropriate to follow the evolution of detail in the solution, implementing
Barrett's ``toolkit''.

Many other important problems require further attention. Primary among
these is the existence of solutions to the sets of equations described
here, both in the time evolution and initial value problems. We also need
to verify the presence of local causality in this method, as outlined
above, and the corresponding existence of a Courant limit.  The detailed
effects of the approximate symmetry, and its optimum exploitation, also
require investigation. These issues are being addressed in current work
(Barrett {\it et al\/} 1993, Galassi and Miller 1993).

I am grateful to I Pinto and J Miller, who organised the very stimulating
{\it Workshop on numerical applications of Regge calculus and related topics}
in Amalfi in 1990, where some of this work was carried out. I am indebted to
the participants at that workshop for discussions, in particular J Barrett, I
Drummond, W Miller, R Sorkin and R Williams.

\bigskip
\leftline{\bf References}

{\frenchspacing\def\ref{\hangafter=1\hangindent=20pt}\parindent=0pt
\smallskip
\ref
Barrett J W 1992 A mathematical approach to numerical relativity in {\it
Proc. Southampton University workshop on approaches to numerical
relativity (1991)} to appear

\ref
Barrett J W, Galassi M, Miller W A, Sorkin R, Tuckey P A and Williams R M 1993
in preparation

\ref
Galassi M and Miller W A 1992 Lapse and shift in Regge calculus {\it Phys.
Rev. \rm D} to appear

Galassi M and Miller W A 1993 in preparation

Hartle J B 1985 {\it J. Math. Phys. \bf 26} 804--14

Hartle J B 1986 {\it J. Math. Phys. \bf 27} 287--95

\ref
Miller W A 1986 in {\it Dynamical spacetimes and numerical relativity} ed J
Centrella (Cambridge: Cambridge University Press) pp 256--303

Piran T and Strominger A 1986 {\it Class. Quantum Grav. \bf 3} 97--102

Porter J D 1987 {\it Class. Quantum Grav. \bf 4} 375--89

Regge T 1961 {\it Nuovo Cimento \bf 19} 558--71

Ro\v cek M and Williams R M 1984 {\it Z. Phys. \rm C \bf 21} 371--81

Sorkin R 1975 {\it Phys. Rev. \rm D \bf 12} 385--96

Tuckey P A 1989 {\it Class. Quantum Grav. \bf 6} 1--21

\ref
Tuckey P A 1991 Issues in discrete-time 3+1 Regge calculus in {\it Proc.
workshop on numerical applications of Regge calculus and related topics
(Amalfi 1990)} ed I Pinto to appear

Williams R M 1990 private communication

Williams R M and Tuckey P A 1992 {\it Class. Quantum Grav. \bf 9} 1409--22
\par}

\bigskip
\leftline{\bf Figure caption}
\smallskip
The 3-dimensional analogue of the basic evolution step described in the
text. We assume no edges on surface $\Sigma_2$ also lie on $\Sigma_1$. All
tetrahedra shown here are new, added onto $\Sigma_2$. The ``diagonal'' edges
$\mu_1\lambda',\ldots,\mu_{N_1}\lambda'$ and the ``vertical'' edge
$\lambda\lambda'$ are new. The ``horizontal'' edges
$\mu_1\lambda,\ldots,\mu_{N_1}\lambda$ and the vertical edge lie on the
interior of the triangulated region, and provide equations of motion to be
solved for the squared lengths of the new edges.

\bye